\documentclass[10pt,journal,compsoc]{IEEEtran}

\usepackage{graphicx}
\usepackage[ruled,linesnumbered]{algorithm2e}
\usepackage{caption}
\usepackage{dblfloatfix}    
\usepackage{amsmath}
\usepackage{booktabs}
\usepackage{tabularx}
\usepackage{xcolor}
\usepackage{color,soul}
\usepackage{diagbox,pict2e}

\newcolumntype{L}{>{\centering\arraybackslash}m{2.5cm}}

\ifCLASSOPTIONcompsoc
  \usepackage[nocompress]{cite}
\else
  \usepackage{cite}
\fi
\ifCLASSINFOpdf
\else
\fi
\begin{document}
%
\title{Side-Channel Attack Analysis on In-Memory Computing Architectures}
%
%
%
%

\author{Ziyu Wang,~\IEEEmembership{Member,~IEEE}
        Fan-hsuan Meng,~\IEEEmembership{}
        Yongmo Park,~\IEEEmembership{Graduate Student Member,~IEEE} \\
        Jason K. Eshraghian,~\IEEEmembership{Member,~IEEE} 
        and Wei D. Lu*,~\IEEEmembership{Fellow,~IEEE}
\IEEEcompsocitemizethanks{\IEEEcompsocthanksitem Z. Wang, F. Meng, Y. Park, J. K. Eshraghian and W. D. Lu are with the Department
of Electrical Engineering and Computer Science, the University of Michigan, Ann Arbor,
MI, 48109, USA.\protect\\
Corresponding Author: Wei D. Lu (wluee@umich.edu)}
\thanks{Manuscript received August 8, 2022; revised December 22, 2022; accepted March 13, 2023.}}

%
%

\markboth{IEEE TRANSACTIONS ON EMERGING TOPICS IN COMPUTING}%
{Shell \MakeLowercase{\textit{Z. Wang et al.}}: Bare Advanced Demo of IEEEtran.cls for IEEE Computer Society Journals}
%



\IEEEtitleabstractindextext{%
\begin{abstract}
  In-memory computing (IMC) systems have great potential for accelerating data-intensive tasks such as deep neural networks (DNNs). As DNN models are generally highly proprietary, the neural network architectures become valuable targets for attacks. In IMC systems, since the whole model is mapped on chip and weight memory read can be restricted, the pre-mapped DNN model acts as a ``black box'' for users. However, the localized and stationary weight and data patterns may subject IMC systems to other attacks. In this paper, we propose a side-channel attack methodology on IMC architectures. We show that it is possible to extract model architectural information from power trace measurements without any prior knowledge of the neural network. We first developed a simulation framework that can emulate the dynamic power traces of the IMC macros. We then performed side-channel leakage analysis to reverse engineer model information such as the stored layer type, layer sequence, output channel/feature size and convolution kernel size from power traces of the IMC macros. Based on the extracted information, full networks can potentially be reconstructed without any knowledge of the neural network. Finally, we discuss potential countermeasures for building IMC systems that offer resistance to these model extraction attack.
\end{abstract}

\begin{IEEEkeywords}
side-channel attack, power analysis, in-memory computing, neural networks, deep learning security
\end{IEEEkeywords}}

\maketitle

\IEEEdisplaynontitleabstractindextext

%
\IEEEpeerreviewmaketitle

\ifCLASSOPTIONcompsoc
\IEEEraisesectionheading{\section{Introduction}\label{sec:introduction}}
\else
\section{Introduction}
\label{sec:introduction}
\fi

\begin{figure}[h]
  \centering
  \includegraphics[width=\linewidth]{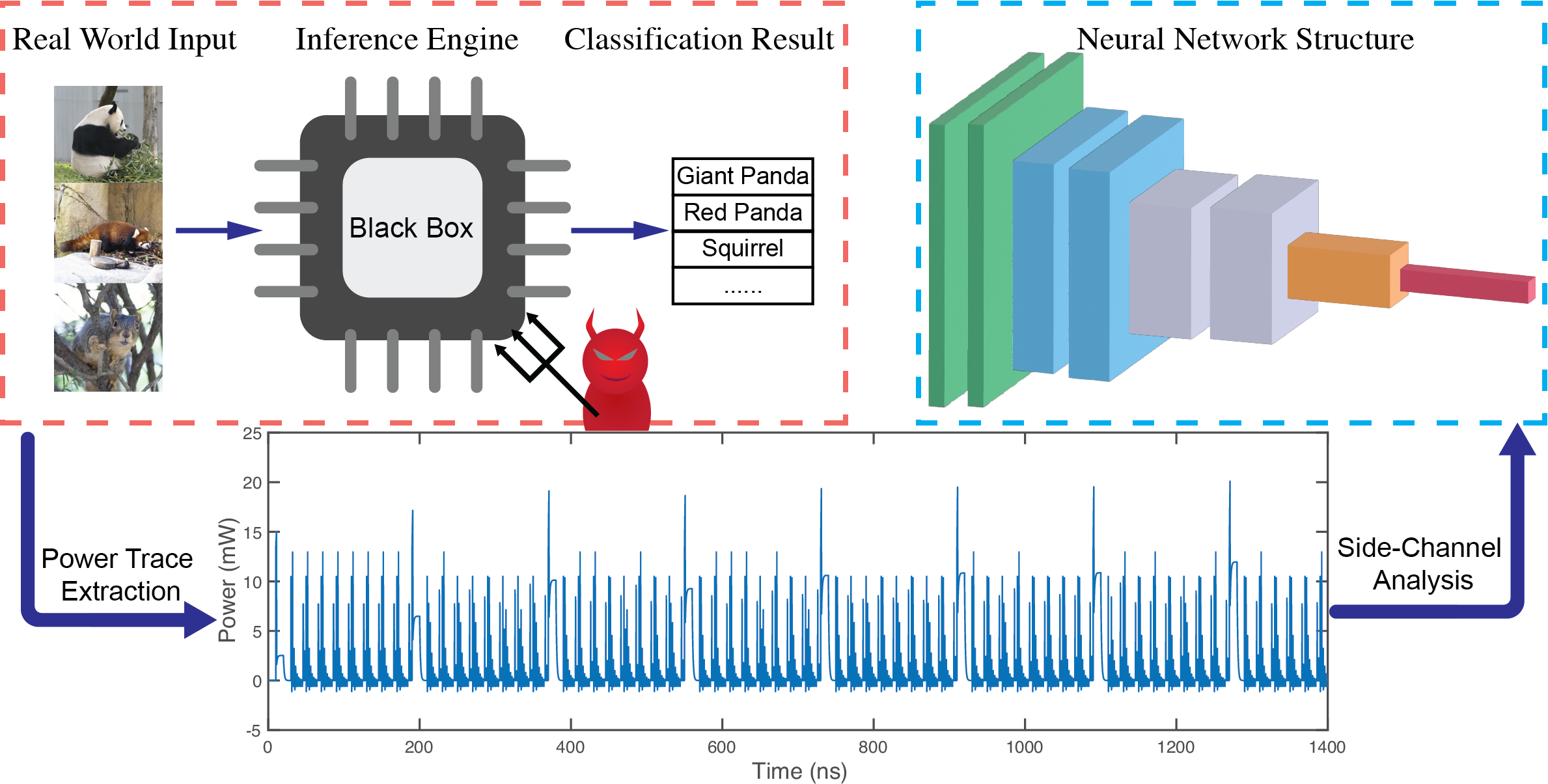}
  \caption{Schematic of side-channel attack on IMC chip.}
  \label{fig1}
\end{figure}

%
%
%
%
In-memory computing (IMC) architectures can circumvent von Neumann's bottleneck when accelerating communication-limited tasks, such as deep learning workloads \cite{chen2020survey} \cite{wang2021rram} \cite{chi2016prime}. However, the security vulnerabilities of analog IMC architectures are yet to be evaluated, and this becomes of paramount importance when the target market of low-power IMC accelerators are in ubiquitous, edge-based computing that transmit and receive information from a variety of sensors \cite{arafin2020security} \cite{khan2021comprehensive}. In theory, side-channel attack can be used to extract and infer information pertaining to the on-chip deep neural network (DNN) model deployed during inference \cite{xiang2020open}, as shown in Figure \ref{fig1}.

\begin{figure}[t]
  \centering
  \includegraphics[width=\linewidth]{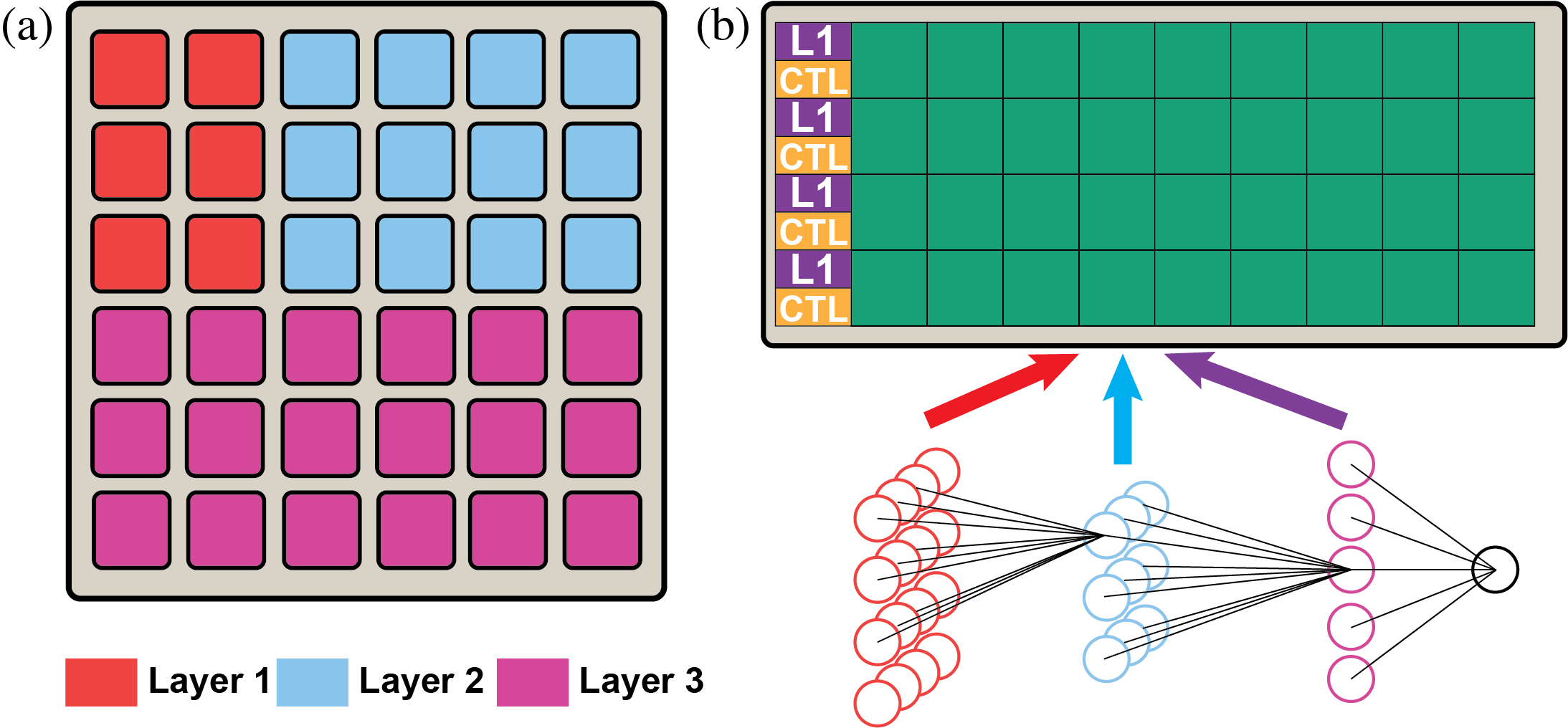}
  \caption{(a) IMC and (b) GPU for DNN acceleration.}
  \label{fig:2}
\end{figure}

When IMC is coupled together with mixed-signal computation, as with Resistive Random-Access Memory (RRAM) crossbar hard macros, it is thought that analog computation via bit line current summation or charge accumulation offers a further layer of obfuscation. Co-locating memory and processing in a tiled architecture eliminates data movement between memory and processor \cite{chi2016prime} \cite{zidan2018future}, as shown in Figure \ref{fig:2}(a). Since the weights are stationary, weight memory access can be further limited. Hence, it is more challenging for malicious users to compromise the security of these hardware accelerators. Conventional DNN accelerators, such as GPUs, use single-instruction-multiple-threads/data (SIMT/SIMD) execution and must therefore time multiplex operations that take place across different DNN layers, as depicted in Figure \ref{fig:2}(b). Rich data-dependent information, such as read/write volume, memory address track, and execution latency, can be obtained from a bus-snoop attack or side-channel attack \cite{hu2020deepsniffer}.

While IMC macros reduce the number of possible attack vectors, the risk of side-channel attack remains a looming threat, as information leaks may still occur via power profiling and electromagnetic emanations. Timing analysis of thread-level execution may also offer an unintended window into architectural insights. Particularly, since different layers of the network are mapped in different IMC macros on chip with fixed data patterns, side-channel attack on the IMC macros may reveal architectural information of the DNN models. The adversary can then counterfeit the intellectual property of the DNN design, or otherwise boost the success rate when conducting multiple attacks to strike the system, such as adversarial attacks \cite{gu2014towards} or adversarial patches \cite{brown2017adversarial}.

In this paper, we demonstrate the complete network architecture for DNN models stored inside the isolated memory blocks in an IMC system may be extracted from the power trace of each tile, without prior knowledge of the model. Our side-channel attack was performed by simulating the dynamic power trace of the mixed-signal RRAM IMC macros. By analyzing the power trace of the different IMC macros, the layer type and sequence, output channel/feature size of convolutional/fully connected layers as well as kernel size of convolutional layers, can be inferred. As an example, we demonstrate how we are able to reverse engineer the full LeNet \cite{lecun1989backpropagation} architecture from a mixed-signal RRAM accelerator by side-channel analysis, without prior additional knowledge of the NN model in use. We also propose several countermeasures that can potentially make the IMC systems resistant to such side-channel attacks, which should be considered during hardware and compiler design.

\section{Background and Motivation}
\label{sec:Background}

\subsection{Analog IMC Architecture}
\begin{figure}[t]
  \centering
  \includegraphics[width=\linewidth]{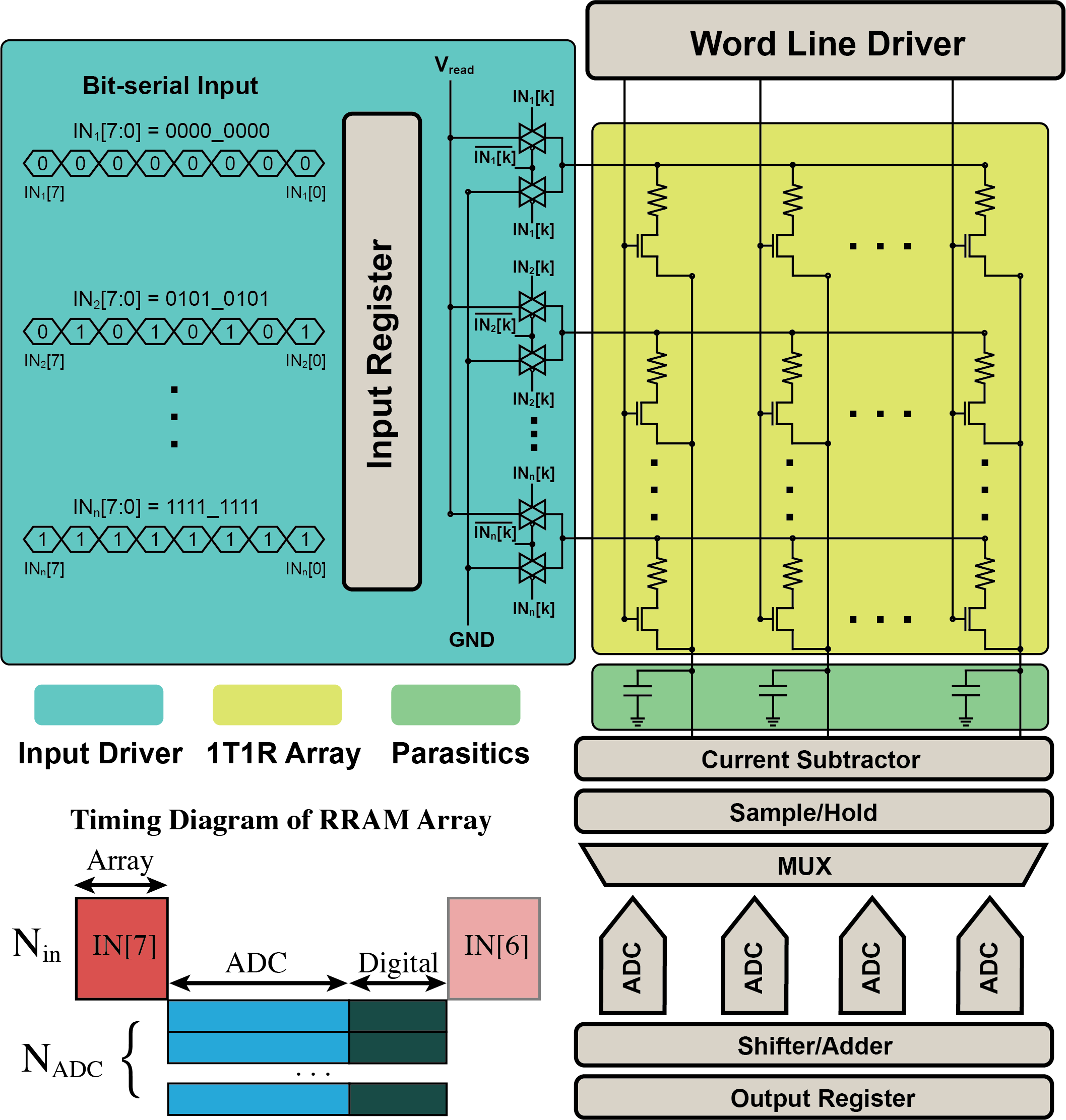}
  \caption{RRAM-based IMC architecture.}
  \label{fig:3} 
\end{figure}

Analog IMC architectures offer significant advantages in throughput and power efficiency by minimizing data movement \cite{chi2016prime} \cite{zidan2018future}, and offer a high degree of parallelism at run time \cite{wang2021rram} \cite{prezioso2015training} \cite{hung2021challenges} with respect to multiply-and-accumulate (MAC) operations.  Figure~\ref{fig:3} shows a detailed schematic of an analog RRAM-based IMC architecture. An RRAM array can perform vector-matrix multiplication (VMM) directly in a single step based on current summation: Ohm's law is used for multiplication and Kirchhoff's current law is for accumulation. More specifically, the input vectors are encoded as voltage pulses and the entries of matrices are mapped as RRAM device conductances. The outputs of the VMM are returned as bit line currents, subsequently sampled by peripheral read-out circuitry, and converted to binary digital values using analog-to-digital converters (ADC) for further downstream communication and processing. As the weight precision is often higher than the device precision, typically a single weight value is mapped across multiple RRAM cells, and the VMM result is reconstructed using digital shifter and adder circuits.

Non-ideal effects such as device variation and IR drop from wire resistance limit the number of rows and columns in an array to typically $128 \times 128$  to perform these analog operations\cite{wang2021device} \cite{zhu2020insights}. Generally, the DNN weight matrices are much larger than the practical analog RRAM IMC macro sizes, and need to be mapped to multiple IMC macros in a tiled architecture through digital interfaces, as shown in Figure \ref{fig:2}(a).

\subsection{IMC Simulators}

\begin{figure*}[!b]
  \centering
  \includegraphics[width=\textwidth]{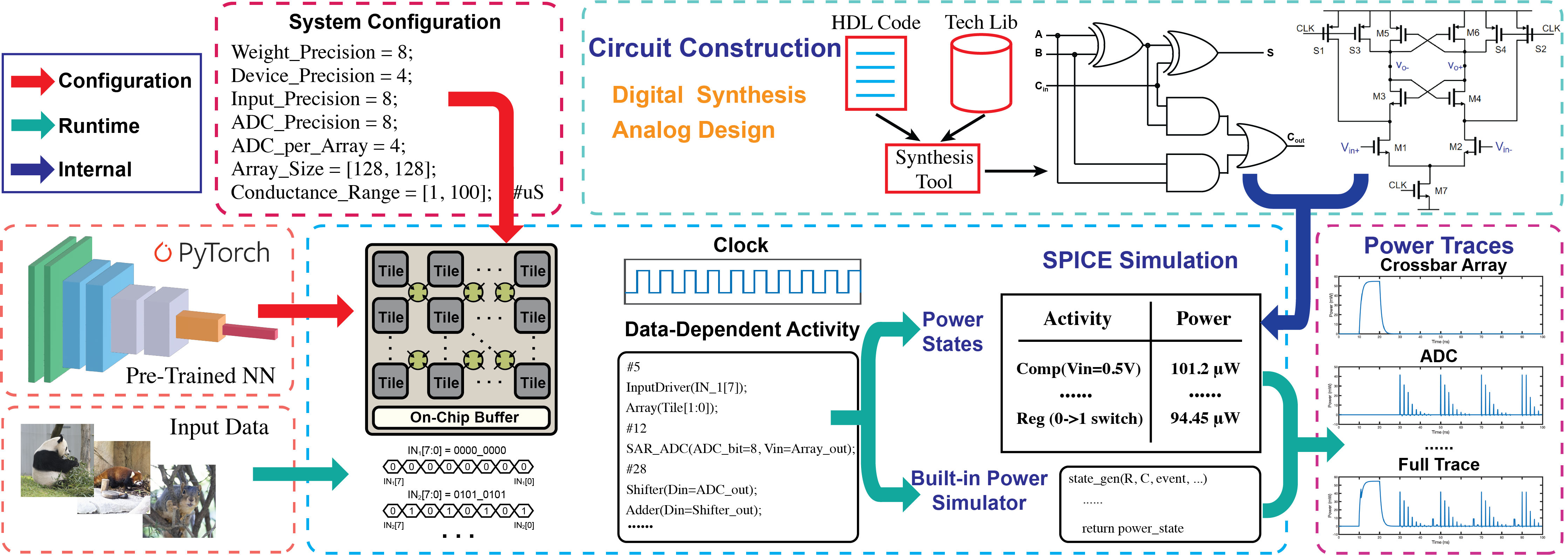}
  \caption{Dynamic power simulator framework.}
  \label{fig:4}
\end{figure*}

Modern IMC applications such as DNNs for machine learning may require a large number of such IMC tiles, making it impractical to perform analog SPICE simulations that can account for all parasitics \cite{zhang2020cccs}. Instead, the hardware performance is often estimated through simulators that integrate simplified device and circuit models to produce inference accuracy, hardware power and area, with rapid iteration times.

Architecture-level simulators such as MNSIM \cite{xia2017mnsim} and NeuroSim \cite{chen2018neurosim} provide a flexible interface for a wide range of design options, and can effectively simulate hardware performance (e.g. power, area, and latency) based on simulator-embedded circuit-level models. However, these works, amongst other simulators \cite{lammie2022memtorch} \cite{rasch2021flexible} \cite{peng2019dnn+} \cite{lin2022dl} \cite{roy2021txsim},  mainly focus on inference accuracy and global hardware performance. Time dependent data, e.g. dynamic power traces, are less explored. As we will demonstrate, dynamic power and other related time-varying information that is available from read-out ports of RRAM macros can prove to be useful in performing side-channel attack, and thus expose an underexplored vulnerability in mixed-signal IMC accelerators. Furthermore, dynamic power information is also helpful for thermal aware optimization \cite{shin2020thermal} \cite{beigi2018thermal}.

\subsection{Related Work and Motivation}

DNN accelerators are increasingly susceptible to malicious model extraction attacks, which expose the network architectural information. Several initial studies on model extraction attacks have been proposed on GPU, CPU and other DNN accelerator platforms \cite{xiang2020open} \cite{hu2020deepsniffer} \cite{hua2018reverse} \cite{yan2020cache} \cite{batina2019csi} \cite{zhang2021stealing}. DeepSniffer \cite{hu2020deepsniffer} and ReverseCNN \cite{hua2018reverse} extract DNN models from GPUs and general DNN accelerators. However, their attack model heavily relies on data movement in the memory bus, and cannot be generalized over to IMC architecture. Model extraction attacks with power or electromagnetic side-channel attack have also been reported on ARM cortex-based systems and FPGAs \cite{xiang2020open} \cite{batina2019csi} \cite{zhang2021stealing}. However, to the best of our knowledge, model extraction attacks have not been reported on analog IMC-based DNN accelerators.

The ability to extract pre-mapped DNN models from IMC chips as discussed in this study highlights a critical vulnerability issue of IMC architectures, and identifying how model information can be reverse-engineered from measurable electrical parameters can provide a guideline for designing more secure and trustworthy IMC accelerators.

\section{Simulation Framework}
\subsection{Overview}
To analyze the dynamic power information of IMC systems, we first developed a mixed-signal power simulator. The overall simulator framework is shown in Figure \ref{fig:4}. The simulator offers two interfaces: i) one during configuration that allows users to define hardware-level properties (system configurations), and perform mapping of a pre-trained NN model through a PyTorch\cite{paszke2019pytorch} interface, and ii) another during runtime simulation, which takes a dataset as an input. The pre-trained weights are mapped to RRAM conductance values based on weight and device precision, as well as the permissible device conductance range, following standard approaches \cite{chen2020survey} \cite{wang2019deep}. Each feature (or pixel) of the input data samples, e.g., images, is scaled and converted to a bit-serial input, across a given number of clock cycles determined by the specified input precision.

The high-level hardware architecture of our simulator is similar to other work mentioned in the previous section, e.g. NeuroSim \cite{chen2018neurosim} \cite{chen2017neurosim+} \cite{lu2021neurosim}. Different from prior efforts that aimed to simulate the hardware performance as a whole, or the performance of each component at runtime, we target the time dependent power data that can be used to reverse engineer the tasks being performed in the IMC systems. The simulator is designed to provide reliable dynamic power information based on the input and memory data patterns, while enabling rapid experimental iterations. In particular, the dynamic switching power is calculated at each clock cycle, since the transition power is data dependent and will be critical for power trace analysis \cite{lorenz2012non}. To provide reliable power simulations, we synthesized the deployed digital components (e.g. adder, register) and custom-designed analog components (e.g. ADC, MUX) using a TSMC 28 nm technology. Complex digital circuit components were modeled by sub-dividing them into basic units that can process 1 bit data, where high-to-low and low-to-high transition powers are extracted based on the technology database. For analog components such as the comparator in the ADC, the dynamic power of each input voltage is recorded at 100 mV intervals. The recorded power data is used to generate a built-in lookup table (LUT). The circuit modules subsequently refer to the LUT and generate their power transition states. For simpler resistive and capacitive circuits, such as RRAM and CDAC in successive approximation registers (SAR) ADC, we developed a built-in power simulator to compute their power traces. Each circuit component will generate its own power trace which is subsequently merged as the full trace for a given tile.

\subsection{RRAM Array and ADC}
Schematic and timing details of the RRAM array model are provided in Figure \ref{fig:3}. During the inference phase, the word line drivers turn on the select transistors in the 1T1R structure. The input data are converted into voltage pulses in bit-serial fashion, and applied to each source line. The currents through the RRAM devices are accumulated along the vertical bit lines. We include the effect of the parasitic capacitance seen from the bit line which introduces a propagation delay during the inference phase. To account for positive and negative weights, we implement a current mirror-based subtractor circuit, which subtracts the output current from the positive and negative weight columns. The subtracted current will charge or discharge a sampling capacitor, which has been pre-charged to VDD/2. The capacitor will hold the analog voltage for the ADC, before the outputs are digitized.

\begin{figure}[h]
  \centering
  \includegraphics[width=\linewidth]{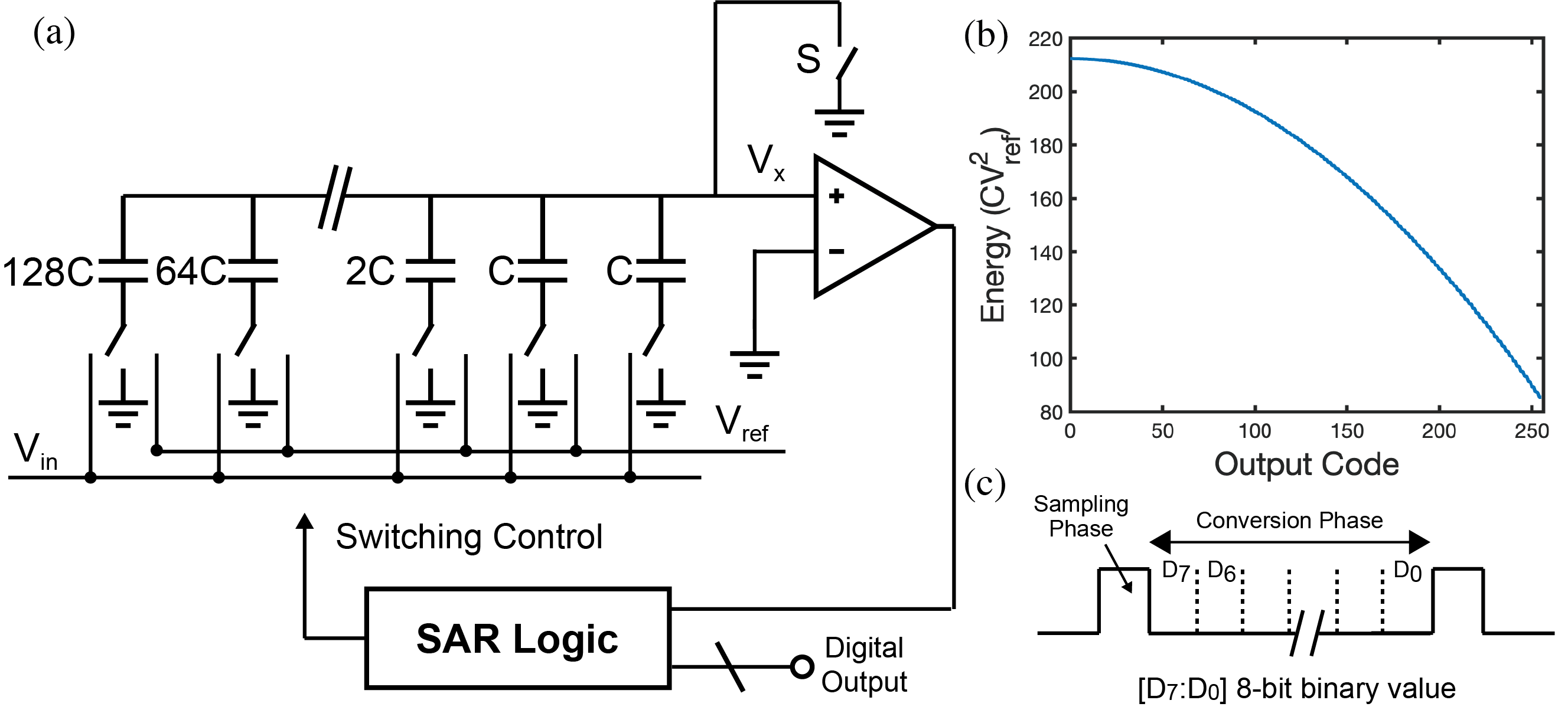}
  \caption{(a) Schematic of an 8-bit SAR ADC. (b) DAC switching
  energy with respect to output codes. (c) Timing diagram of the SAR ADC.}
  \label{fig:5}
\end{figure}

Figure \ref{fig:5}(a) shows the schematic of a synthesized 8-bit charge redistribution SAR ADC design in the simulation, and Figure \ref{fig:5}(c) depicts its timing diagram. During the sampling phase, the switch $S$ is closed and all capacitors are connected to $V_{in}$. Next, the switch $S$ is opened and all capacitors are grounded via their bottom terminals, raising the voltage at the positive input terminal of the comparator to $-V_{in}$. During the conversion phase, SAR logic controls the switches to $V_{ref}$ one by one to perform a binary search. The conversion takes $8$ cycles and the energy consumption of the $n$-th step can be calculated from the change in $V_x$, and the capacitance connected to $V_{ref}$ using Equation \ref{eq:1}.

\begin{equation}
   E=
   \begin{cases}
   -C_{1}V_{ref}(\Delta V_{x} - V_{ref}), & n = 1\\ 
   -V_{ref}(\Delta V_{x} \sum_{i=1}^{n-1}C_{i}D_{i} + C_{n}(\Delta V_{x} - V_{ref})), & n \neq 1
   \end{cases}
  \label{eq:1}
\end{equation}

\noindent where $D_{i}$ is the $i$~th MSB output code. When $D_{i} = 1$, the $i$-th switch connects to $V_{ref}$, otherwise, it connects to ground.

The dynamic power from the transitions in the DAC is the dominant contribution to the total power consumed by the SAR ADC \cite{ginsburg2005energy} \cite{hariprasath2010merged}. Figure \ref{fig:5}(b) shows the switching energy with respect to the output code, and it has a clear data pattern dependency. In our simulator setup, the unit capacitor in the DAC array is set as 1~fF, and each capacitor can be charged and settled within a worst-case interval of 20~ps.

\subsection{Digital Components}
The digital components of an IMC system include input and output registers, shifters and adders within a tile, as well as inter layer logic functions such as ReLU activations, pooling operators, and routing circuits. As mentioned above, the transition power of digital components are based on a built-in LUT which stores the extracted transition power data of the basic digital units from the TSMC 28 nm technology process. To fully simulate the data flow in the hardware, all data movement between circuit modules in our simulator are performed in a binary, time-multiplexed fashion. Hence, the total dynamic power of each switching event from all digital components can be calculated based on each input bit of data.

\subsection{Model Mapping}
\begin{figure}[h]
  \centering
  \includegraphics[width=\linewidth]{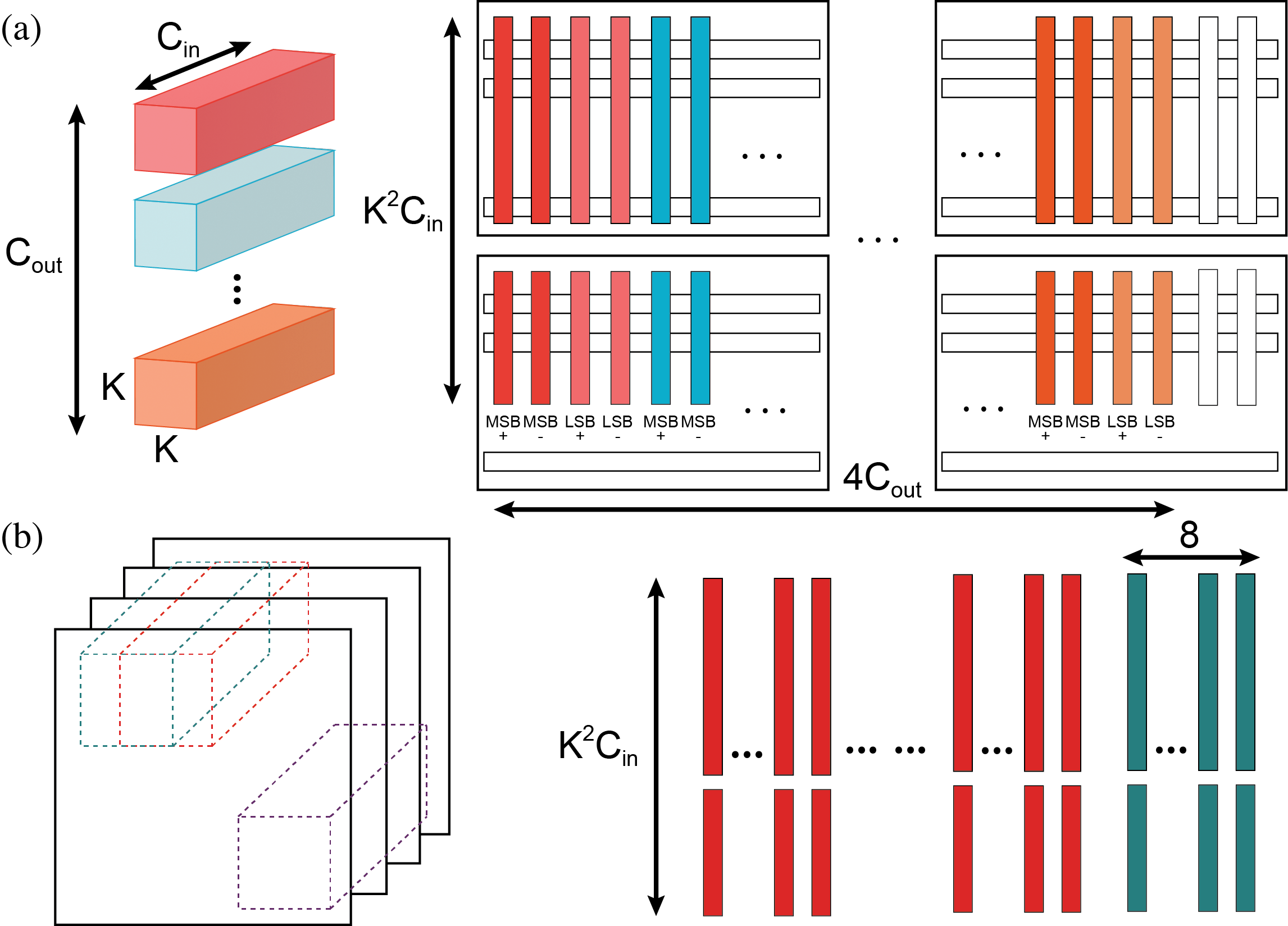}
  \caption{(a) Convolution kernel mapping. (b) Input data mapping.}
  \label{fig:6}
\end{figure}

For inference tasks, the pre-trained weights are mapped across tiled RRAM arrays, and remain stationary during operations. To benefit from inference efficiency, the weights and input activations are quantized to 8 bits. It has been demonstrated that 8-bit weight precision does not incur a significant accuracy degradation, especially when coupled with quantization-aware training techniques \cite{jacob2018quantization} \cite{eshraghian2022navigating}. However, RRAM cells do not have sufficient internal precision to support 8-bit weights, which requires reliable programming of 256 conductance levels. As a result, it is often more practical to map a given weight across multiple cells, e.g. using 2 cells each offering 4 bits. Figure \ref{fig:6}(a) shows the mapping approach used in our simulation for the convolution kernels. Other mapping strategies can be employed similarly through the configuration interface shown in Figure \ref{fig:4}. Each kernel is flattened to a 1D vector, followed by splitting positive and negative values across columns. Next, each weight is quantized to 8 bits, mapped between the 4 MSBs and 4 LSBs separately. For an IMC system with 8-bit weight precision and 4-bit device precision, the total column number is thus 4 times that of the output channel size (2 for positive and negative weights, and 2 for splitting the 8-bit weights into 2 cells), and the number of mapped rows is the flattened kernel size $K^2C_{in}$, where $K$ is a single kernel dimension and $C_{in}$ is the input channel size.

The output of the convolutional layer is computed by sliding the input activation across the kernels, as illustrated in Figure \ref{fig:6}(b). The input to each kernel is flattened to match the kernel, and outputs from all kernels are computed simultaneously through the IMC module outputs. For each 8-bit input, we employ bit-serial representation and it takes 8 consecutive steps to compute 1 input.

The weights and inputs to a fully connected layer can be quantized and mapped directly without any reshaping. Similarly, vector outputs of the fully connected layer can be acquired simultaneously through the IMC outputs.

\section{Experiment and Analysis}
\subsection{Experimental Setup}
After model mapping, we simulate the DNN inference in the IMC system at the circuit-level, and use the extracted power traces to analyze the feasibility of model extraction attacks through the side-channel leakage. 
Side-channel attack analysis is based on physical phenomena during execution, as well as mathematical analysis. By carefully measuring and analyzing the power dissipation of the chip, attackers may be able to reverse engineer sensitive data or architectural information \cite{kocher1996timing}. A general assumption for the proposed attack is that the attacker already knows hardware parameters of the IMC tiles, i.e., the RRAM array size, the ADC type and the number of ADCs per tile, but has no knowledge about the mapped NN. This is a realistic assumption since memory access to the stored weights in IMC systems can be restricted and physically separated from other programs the attacker may gain access to. The attacker, as a user, however has access to the input and output ports of the chip, i.e., can control the input data supplied to the IMC system during runtime.
For example, most of inference tasks of RRAM-based IMC system are pipelined during processing to improve the throughput\cite{chi2016prime} \cite{wang2019deep} \cite{song2017pipelayer}, i.e., starting to process the second input as the first input passes through the first layer. This will blur valuable power and timing information of a single runtime. However, as the attacker has the full control of the input and output data ports, she/he can halt the next input to the system until the previous inference completion to gain more accurate power measurements.

\begin{figure}[h]
  \centering
  \includegraphics[width=\linewidth]{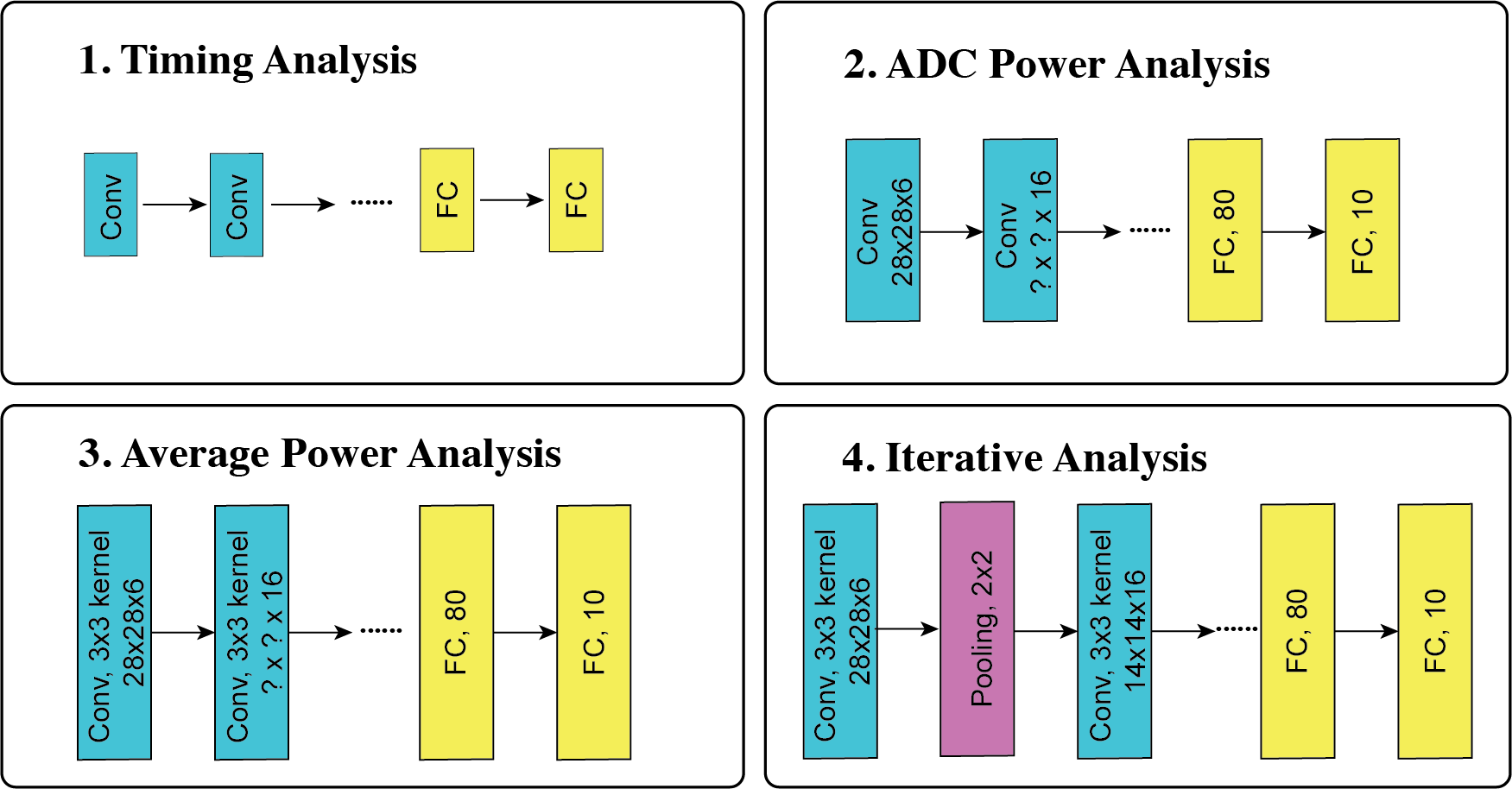}
  \caption{Overview of side-channel attack flow for DNN model extraction.}
  \label{fig:7}
\end{figure}

Side-channel attack can be classified into invasive and non-invasive attacks \cite{fan2010state} \cite{hutle2015resilience}, based on whether decapsulation of the chip is used. Carefully designed invasive attacks make it possible to measure the power trace of each single RRAM tile on chip. In our experiment, the assumption is the attacker can only measure the power traces of each tile for side-channel attack, and does not have access to individual RRAM devices. Figure \ref{fig:7} illustrates a simplified flow for model extraction attack, which will be discussed in the following sections.

We first initialize the hardware design with the system configuration parameters shown in Figure \ref{fig:4}.  The conductance range of the RRAM devices is set between 1~$\mu S$ and 100~$\mu S$. Each cell has 16 conductance levels. A 50 MHz clock is used to drive the bit-serial inputs, limited by the RRAM read operation speed. The other digital logic is assumed to operate at 500 MHz. Before inference, the pre-trained NN model is mapped to RRAM arrays through the interface between the simulator and PyTorch, without any sensitive information leakage. The model is deemed `inaccessible' to a user once the model is compiled onto the IMC accelerator, for the entirety of the side-channel attack. At the very end of the experiment, we validate the extracted model structure through side-channel attack with the ground truth.

In this experiment we use CIFAR-10 \cite{krizhevsky2009learning} as the dataset during inference, which consists of a set of 60,000 natural color images that are 32$\times$32 in size.

\begin{figure*}
  \centering
  \includegraphics[width=\textwidth]{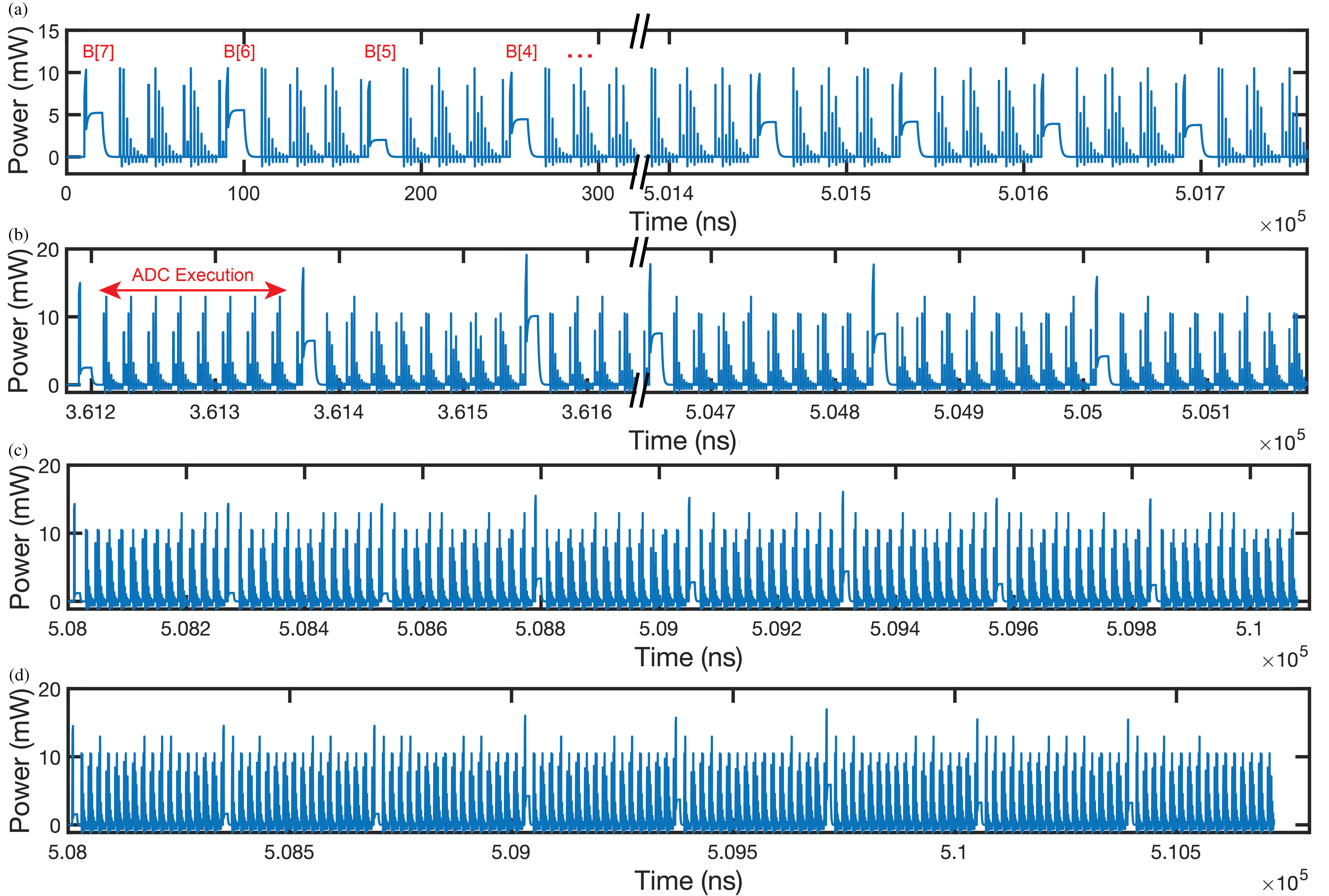}
  \caption{Simulated output power traces from 4 different IMC tiles during one inference task.}
  \label{fig:8}
\end{figure*}

Before elaborating attack approach, we summarized the assumption for the attacker's knowledge as follows.
\begin{itemize}
    \item The attacker knows the hardware implementation details of an IMC tile, including array size, ADC type and number of ADCs per tile. However, the attacker has no knowledge of the neural network mapped on chip.
    \item The attacker has the full control of the input and output ports of the chip. Hence, the next input can be halted until the completion of the previous input, allowing the attacker to control the input pipeline to get more accurate power trace of each inference.
    \item The attacker has no access to individual RRAM cells, but can measure the power trace of each tile as a whole.
\end{itemize}

\subsection{Power Traces}
We test the side-channel attack on the IMC system that has an unknown NN already mapped on chip. Based on the power traces extracted from the simulator, it is found that 23 IMC tiles are utilized to store the pre-trained model. Figure \ref{fig:8} displays the power traces of 4 of those tiles during an inference. Examining the power traces, we can first identify when a bit-serial input is applied to the RRAM array. As can be seen in Figure \ref{fig:8}, processing a 1-bit input in the IMC module is accompanied by the characteristically transient IR power draw from the array, followed by a stable period during data passing the RRAM cells, and a switching-dense period for data conversion in the ADC along with the execution of other digital components. By inspecting the power traces, the following hardware hints can be extracted. 

\begin{itemize}
    \item \textbf{Start time:} corresponding to when a tile starts execution. Tiles with identical start times are expected to belong to the same NN layer (e.g. Figure \ref{fig:8}(c) and (d)). Grouping tiles with the same start time will provide information on the layer size and the number of layers that perform VMMs.
    \item \textbf{Execution time:} corresponding to how many bit-serial inputs are executed at the layer for a given inference. The number of input bits sent to a module can be inferred by identifying the arrival of a new input bit, which has the distinguishing feature of the transient IR power draw followed by a stable period of a few nanoseconds corresponding to data passing through the RRAM cells. Figure \ref{fig:8}(a) illustrates this by including labels for the bit-serial inputs of the first data sample. The execution time of convolutional (Conv) layers are typically much longer than that of fully connected (FC) layers due to the larger number of computational cycles required in convolutions.
    \item \textbf{ADC execution time:} corresponding to how long it takes for the ADC to convert all analog outputs from the RRAM array. ADC execution time can be extracted by inspecting the dynamic power consumption of the ADC after data has passed the RRAM cells. During ADC conversion, the successive approximation steps lead to decrementing voltage changes, and the overall tendency of ADC transition power during a single conversion is also decreasing. The ADC execution time indicates how many columns are being utilized in an array.
    \item \textbf{Average Power:} $\frac{1}{T}\int_{T_1}^{T_2}P(t)dt$. For arrays with the same number of utilized columns, the ratio of average power is statistically proportional to the number utilized RRAM rows.
\end{itemize}

Using these features, we will show how to extract the full NN architecture in the following section.

\section{Neural Network Extraction}
\subsection{Layer Property Extraction}
The number of NN layers that perform VMM operations and their corresponding layer types can be identified first, based on the start time and execution duration. The extraction process of these layer properties is summarized in Algorithm \ref{alg:one}. The algorithm first extracts the start time and execution time of each tile. Whether a tile belongs to a Conv layer or a FC layer can then be identified based on how many VMM operations are executed. The number of VMM operations for a Conv layer corresponds to the output feature map size, but is fixed at `1' for a FC layer. Finally, tiles with the same start time are grouped, and the layer sequence is generated by sorting the start time of these grouped tiles.

In our experimental test case, using this approach, the 23 tiles utilized in the network inference were found to belong to 2 Conv layers and 3 FC layers.


\begin{algorithm}
\caption{Layer Property Extraction}\label{alg:one}
\KwData{Power traces of all tiles}
\KwResult{A list of NN layers that perform VMM operations}
\For{$i = 1$ to TileNum}{
    Tile$[i]$.StartTime $\gets$ StartTimeExtract(trace$[i]$)\;
    Tile$[i]$.VMMExTime $\gets$ ExTimeExtract(trace$[i]$)\;
    \eIf{Tile$[i]$.VMMExTime $== 1$}{
        Tile$[i]$.Type = fc\;
    }
    {
        Tile$[i]$.Type = conv\;
    }
}
    LayerSequence = GroupNSortTiles(Tile)\;
\end{algorithm}

\subsection{Output Channel Size Extraction}
Based on the mapping approach in Section 3.4, we know that the output channel size of a given Conv layer and the output feature size of a FC layer are directly related to the number of the utilized RRAM columns in the IMC modules. The number of utilized columns in each module can be extracted from the ADC execution time. In the system design we analyzed, each tile has 128 columns and generates 64 output currents, where the subtraction of positive and negative bit line currents is performed in the analog domain. Each tile also has 4 ADCs. Hence, each of the 4 ADCs will execute 16 times to convert the 64 outputs for a fully mapped array. In a partially mapped tile, shorter ADC execution time will be observed, so the ADC execution time can be utilized to estimate the output channel size. The process is summarized in Algorithm \ref{alg:two}. We first extract the ADC execution time from the power traces. If the ADC execution time is smaller than the maximum execution time, which is 16 in our case, it will be labeled as a column-wise partially mapped tile (i.e., the array is not fully utilized). The number of partially mapped tiles and fully mapped tiles are counted, which then allows us to calculate how many columns are mapped and the output channel size for the given layer.

\begin{figure}[h]
  \centering
  \includegraphics[width=\linewidth]{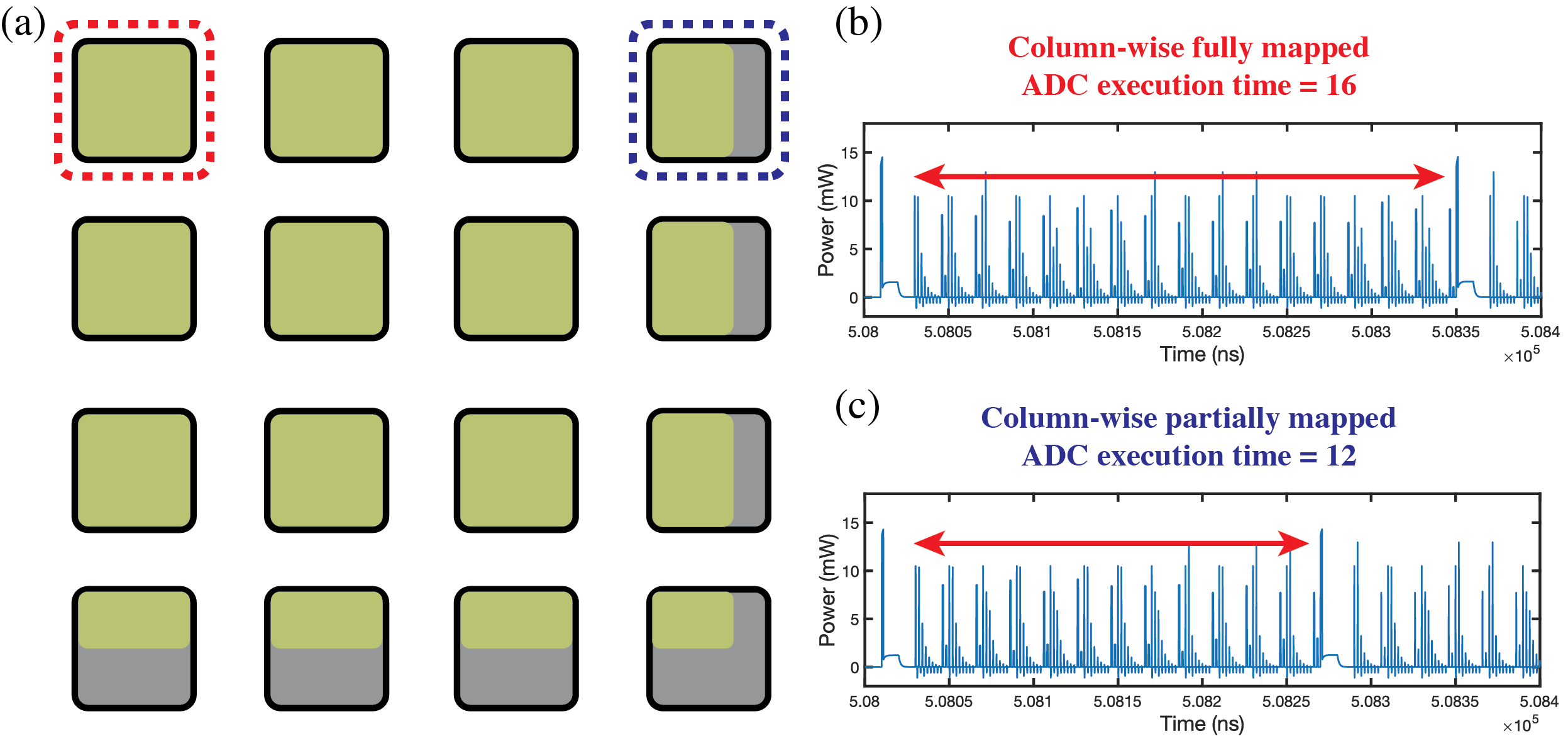}
  \caption{(a) IMC tiles mapped with weights from the first FC layer. 
  (b) ADC execution time of a column-wise fully mapped tile, and (c) a 
  column-wise partially mapped tile.}
  \label{fig:9}
\end{figure}

As an example, Figure \ref{fig:9} plots the extracted utilization and power traces identified from the first FC layer in the unknown NN model. From Algorithm \ref{alg:two}, we can conclude there are 4 column-wise partially mapped tiles. The mapping information of the 16 tiles used to map the FC layer is shown in Figure \ref{fig:9}(a). Figures \ref{fig:9}(b) and (c) display the ADC execution time traces for the two tiles marked in (a) for a given input. From inspection, the ADCs in the tile marked by the red line executed 16 times, indicating it is fully mapped and utilized. The ADCs from the tile marked by the blue line only executed 12 times, indicating 96 columns are mapped. By analyzing the power traces from the 16 tiles, we can calculate the output feature size to be $(3 \times 128 + 96) / 4 = 120$.

\begin{algorithm}
\caption{Output Channel Size Extraction}\label{alg:two}
\KwData{Power traces of tiles in one layer}
\KwResult{Output channel/feature size of Conv/FC layer}
\For{$i = 1$ to TileNum}{
    Tile$[i]$.ADCExTime $\gets$ ADCExTimeExtract(trace$[i]$)\;
    \If{Tile$[i]$.ADCExTime $<$ ColNum/($2 \times$ ADCNum)}{
        ColPartialMapTile$++$\;
        ColNum = Tile$[i]$.ADCExTime$ \times 2 \times $ADCNum\;
    }
}

FullCol $=$ TileNum $/$ ColPartialMapTile - 1\;
OutSize = (FullCol $\times$ ArrayCol $+$ColNum)$/$ CellPerWeight\;
\end{algorithm}

Using the same approach, the output channel sizes of the first two identified Conv layers can be uncovered and found to be 6 and 16 from power traces in Figure \ref{fig:8} (a) and (b). The output feature size of the second identified FC layer is 84. The output feature of the third identified FC layer corresponds to the number of classification classes, and is 10 for the CIFAR10 dataset.

\subsection{Kernel Size Extraction}
The weights of the Conv kernels are mapped across $K^2C_{in}$ rows. Therefore, there are two possible cases: $K^2C_{in}$ is either smaller than (or equal to) the number of rows of a single RRAM tile, or it exceeds the number of rows available and must be distributed across multiple tiles. The number of input channels characteristically increases with deeper Conv layers, so the first case of a kernel fitting within a single array typically only happens with the first layer. 

For the first Conv layer, the size of the input image is known to be $W_{in} \times W_{in}$. As the Conv layer computes one output pixel with one input vector, the output feature map size of the first Conv layer can be speculated from the VMM execution time in Algorithm \ref{alg:one}. Furthermore, the input and output feature maps follow the relationship shown in Equation \ref{eq:2}:

\begin{equation}
    W_{out} = \frac{W_{in} - K + 2P}{S} + 1,
    \label{eq:2}
\end{equation}

\noindent where $W_{out}$ and $W_{in}$ are the output and input widths (equal to the heights), $K$ is the kernel size, $P$ and $S$ are for padding and striding.

The first identified Conv layer was fully mapped within an array. Since the total row number is $K^2C_{in} < 128$, where $C_{in}$ is 3 for RGB images, and $K^2$ must be a square number, only a few kernel sizes are possible. Thus, we can uncover the kernel size of the first Conv layer by testing the range of possible kernel sizes iteratively using Equation \ref{eq:2}. Most kernels use odd-numbered dimensions which further narrows the search space to $K=1, 3, 5$. By testing all cases and comparing with the output shape, we found the kernel size is $5 \times 5$. This approach is also helpful when analyzing the pooling layer, which will be discussed in further detail in the next subsection.

\begin{algorithm}
\caption{Kernel Size Extraction}\label{alg:three}
\KwData{Power traces and input channel size of a Conv layer}
\KwResult{Kernel size of Conv layer}
\For{$i = 1$ to TileNum}{
    Tile$[i]$.AvePower $\gets$ MeanPowerExtract(trace$[i]$)\;
    }
TotalAvePower $\gets$ TileColWiseSum(Tile.AvePower)\;
\If{TileRow $\neq 1$}{
    RefPower = Mean(TotalAvePower[Row$=1:n-1$])\;
}
    TotalRow = (TileRow $-1 +\frac{TotalAvePower[n]}{RefPower})\times$ ArrayRow\;
    KernelSize = round(sqrt(TotalRow$/$InChannel))\;

\end{algorithm}

\begin{figure}[h]
  \centering
  \includegraphics[width=\linewidth]{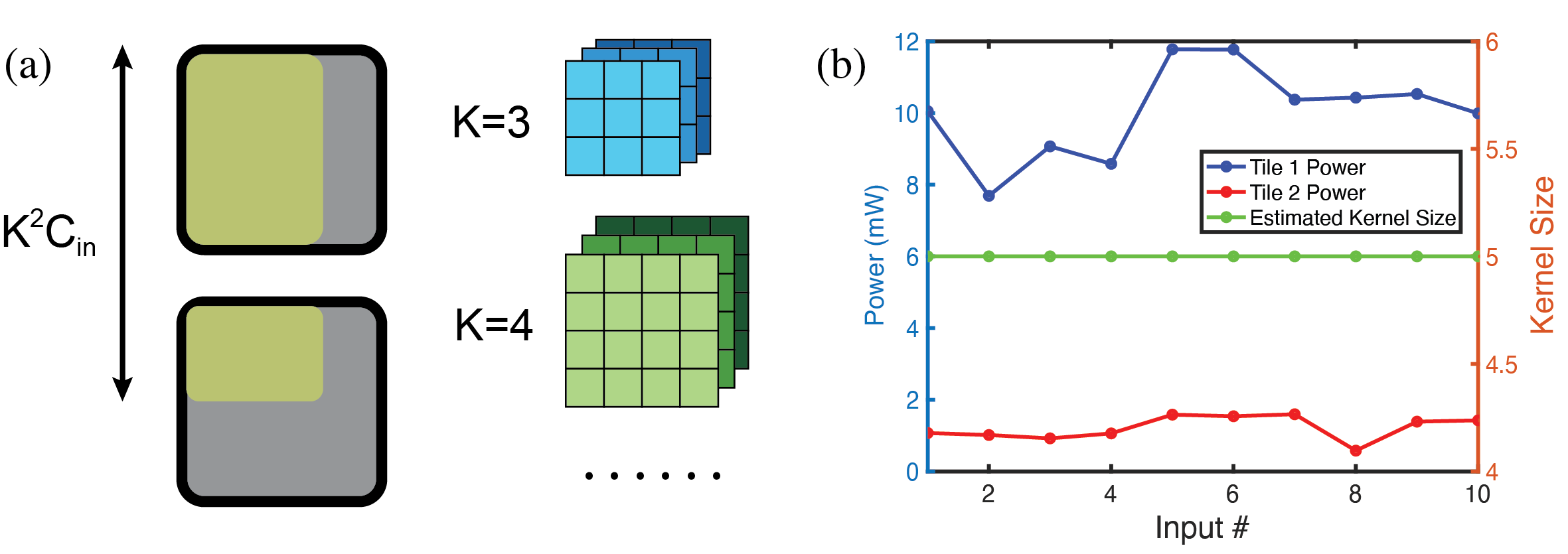}
  \caption{(a) Tiles mapped with weights from the identified second Conv layer and possible kernel sizes. (b) Average power and predicted kernel sizes with respect to different input images.}
  \label{fig:10}
\end{figure}

In most cases beyond the first Conv layer, the input activation includes many channels, and the kernel must be mapped across multiple tiles. Figure \ref{fig:10}(a) shows how the identified second Conv layer utilizes 2 RRAM tiles. For deeper kernels, the kernel size can be extracted by analyzing the average power consumption of all tiles in the Conv layer, with extraction steps summarized in Algorithm \ref{alg:three}. We first extract the average RRAM array power of each tile from the obtained power traces. Next, if the columns are mapped across multiple tiles like in Figure \ref{fig:10}, the average power of each tile row is summed. The reference power is generated by calculating the average power of row-wise fully mapped tiles. The number of utilized rows is obtained from the ratio of the average power of row-wise partially and fully mapped tiles. With this approach, it is not feasible to extract exactly how many rows are mapped. However, as discussed earlier, the total row number is $K^2C_{in}$, and $C_{in}$ has been extracted by Algorithm \ref{alg:two} from the previous layer while $K^2$ must be a square number.  Hence, we only need to find the nearest possible square number that matches the row number estimated from Algorithm \ref{alg:three}. Figure \ref{fig:10}(b) shows the average array power of two tiles in the identified second Conv layer with respect to the input image number. While the average power varies among different inputs, the algorithm always predicts the correct kernel size due to the limited number of permissible kernel sizes.

Using the same approach, the input feature size of the first FC layer can be extracted from the input data dimension which takes the form $N^2C_{out}$, where $C_{out}$ is the number of output channels in the previous Conv layer and has been extracted by Algorithm \ref{alg:two}.

\subsection{Pooling Layer Analysis}
A 2D pooling layer is typically square-shaped, and placed between two Conv layers, or between a Conv and a FC layer. For the latter case, the pooling layer can be easily extracted since the output shape from the Conv layer and the input shape for the FC layer are both known. The ratio of these two indicates how the shape shrinks after the pooling layer and the size of the pooling operation.

However, the information of the pooling layer between two Conv layers cannot be trivially extracted as the input shape of the Conv layer is unknown. Therefore, we propose an iterative search approach summarized in Algorithm \ref{alg:four}. With the output kernel size of the second Conv layer, we can reconstruct a series of input shapes with different padding and stride properties. As the output shape of the first Conv layer is known, we can estimate the output shape of the pooling layer based on different pooling shapes, which is then compared with the reconstructed input shape for the second Conv layer. Table \ref{tab:1} summarizes possible input shapes. Among all of them, only dimensions $14 \times 14$ and $28 \times 28$ can be supported with $2 \times 2$ pooling or without pooling, respectively. The pooling layer is implemented in digital logic, and will incur a delay in the execution start time of the next Conv layer. Hence, based on the above analysis and the delay observed in the start times, we can speculate that a $2 \times 2$ pooling layer exists between two Conv layers.

\begin{algorithm}
\caption{Pooling Detection}\label{alg:four}
\KwData{Output shape and number of channels of the current Conv layer. Output and kernel size of the next Conv layer, or input feature size of next FC layer.}
\KwResult{Size of Pooling layer}
\uIf{NextLayer == fc}{
    PoolingSize = ConvOutShape $/$ sqrt(FcInFeature/ConvOutChannel)\;
}
\ElseIf{NextLayer == conv}{
    InSize $\gets$ TryPadStride(Kernel$\rightarrow$Next, OutSize$\rightarrow$Next) \;
    PoolingSize $\gets$ TryPooling(InSize, OutSize $\rightarrow$ This) \;
}

\end{algorithm}

\begin{table}[h]
  \centering
  \caption{Pooling Layer Detection Attempt}
  \label{tab:1}
  \begin{tabular}{cccl}
    \toprule
    Padding & Stride & Input Size & Pooling\\
    \midrule
    $0$ & $1$ & $14 \times 14$ & \textbf{2 $\times$ 2} \\
    $1$ & $1$ & $12 \times 12$ & - \\
    $2$ & $1$ & $10 \times 10$ & - \\
    $\cdots$ & $\cdots$ &$\cdots$ & $\cdots$ \\
    $2$ & $3$ & $28 \times 28$ & \textbf{1 $\times$ 1} \\
    $3$ & $3$ & $26 \times 26$ & - \\
    $4$ & $3$ & $24 \times 24$ & - \\
  \bottomrule
\end{tabular}
\end{table}

\subsection{Discussion}
\begin{figure}[h]
  \centering
  \includegraphics[width=\linewidth]{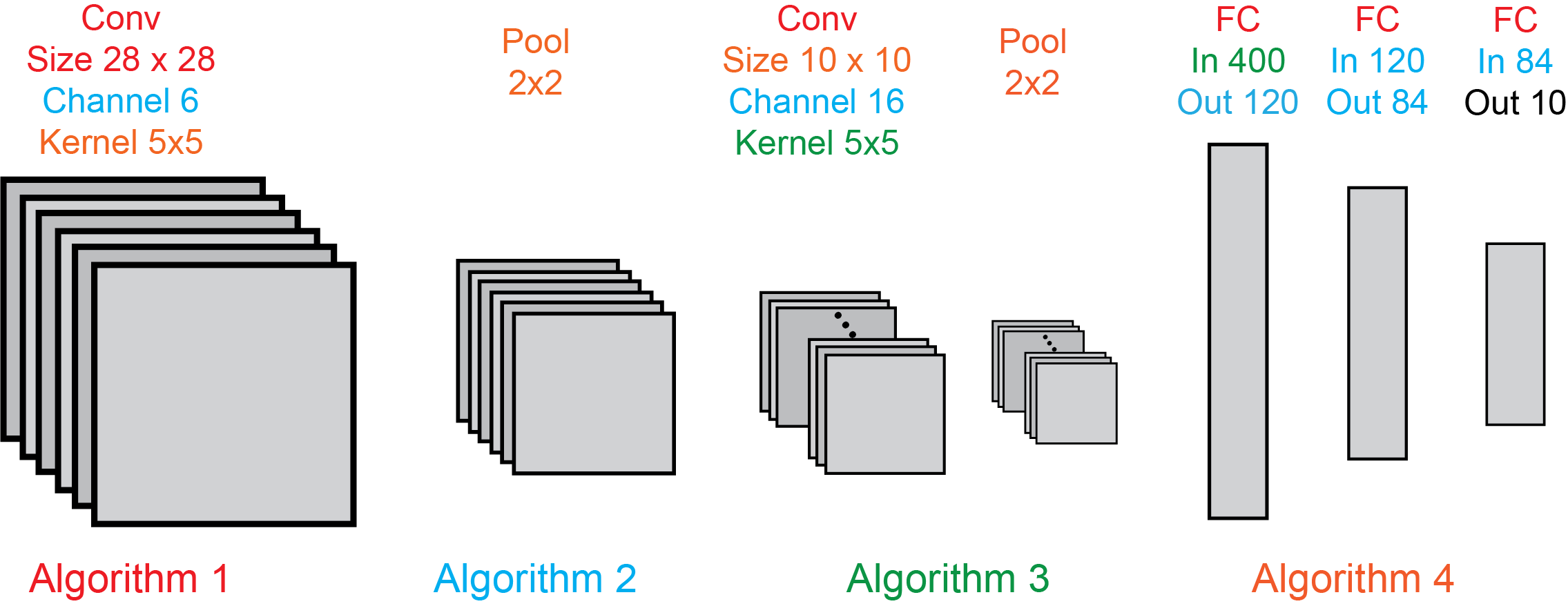}
  \caption{Extracted complete NN architecture for the unknown model. The colors label the algorithm used to obtain the architecture information.}
  \label{fig:11}
\end{figure}

Figure \ref{fig:11} shows the extracted NN architecture based on the analysis above. The colored labels indicate the architectural information that has been extracted from which proposed algorithm. The extracted NN architecture matches the ground truth - a LeNet model, using only side-channel attack without any prior knowledge of the model. The model comparison is further validated with the original PyTorch model file. We expect this approach can be generalised to any NN model that consists of Conv, FC, and 2D pooling layers. Key results of this work and comparison with prior studies are listed in Table \ref{tab:2}.

\begin{table*}[t]
  \centering
  \caption{Comparison of different works performing side-channel attack on DNN accelerators.}
  \label{tab:2}
  \begin{tabular}{l|L|L|L|L|L}
    \hline
     & Platform & Attacking Target & Leaked Data & Data Acquisition & Extraction result \\ \hline \hline
     DAC '18 \cite{hua2018reverse} & FPGA accelerator + off-chip memory & DNN model & Memory and timing & Simulation & Number and type of layers, input/output sizes of each layer, size of filters, weights\\ \hline
     ACSAC '18 \cite{wei2018know} & FPGA accelerator & First layer of CNN and input image & Power & Oscilloscope measurement & Binary input images \\ \hline
     USENIX Security '19 \cite{batina2019csi} & ARM Cortex-M3 microcontroller & DNN model & electromagnetic emanation and timing & Electromagnetic probe measurement & Number and type of layers, input/output sizes of each layer, activation functions, weights \\ \hline
     ASPLOS '20 \cite{hu2020deepsniffer} & GPU & DNN model & Memory access volume and timing & Monitoring PCIe and GDDR memory bus & Number and type of layers, layer topology connection, input/output sizes of each layer, size of filters, activation functions \\ \hline
     TCSII \cite{xiang2020open} & Raspberry Pi & DNN model & Power & External power data acquisition card & DNN architecture type, parameter sparsity \\ \hline
     TIFS \cite{zhang2021stealing} & FPGA accelerator & DNN model & Power & Ring Oscillator Power sensor & Number and type of layers, input/output sizes of each layer, size of filters, activation functions \\ \hline
     \textbf{This Work} & RRAM-based analog IMC accelerator & DNN model & Power and timing & Simulation & Number and type of layers, input/output sizes of each layer, size of filters \\ \hline
\end{tabular}
\end{table*}

\section{Feasibility Analysis}

The above-mentioned side-channel attack methodology is proven capable to reverse engineer the unknown neural network architecture from the power traces. To validate our approach, real-world constraints need to be considered. In this section, we discuss potential data acquisition solutions, sampling rate requirements and circuit noise effects when attacking a real IMC system using the proposed theoretical methodology.

Two measurement techniques can be considered for these data dependent signals. One technique is using electromagnetic probes. As RRAM tiles with different weights are spatially located on the chip, electromagnetic probes with sufficient spatial resolution can measure the side-channel leakage of individual tiles. Tile areas of RRAM-based IMC prototype chips vary due to different array sizes and peripheral circuit designs, but most demonstrated tile areas are on the millimeter scale \cite{wan2022compute} \cite{li202240nm} \cite{correll20228} \cite{yi2022activity}. For example, the reported overall area of a tile with $128 \times 128$ array size, same as our system configuration, is $0.5 \times 0.5~mm^2$ \cite{li202240nm}. The spatial resolution of electromagnetic probes can achieve sub-millimeter \cite{funato2006magnetic} \cite{chou2013space} \cite{peng2019pair}, to make it possible to measure the leakage signal of a tile with no overlapping of adjacent tiles. The second technique is measuring the signal through the power lines of each tile directly. Such techniques have been experimentally used to attack a spin-transfer torque MRAM system for Advanced Encryption Standard (AES) executions  \cite{khan2021comprehensive} \cite{khan2017side}. Like MRAM systems, RRAM devices are fabricated in the back-end-of-line so etching away the passivation layers can potentially provide immediate access to top-level metal lines for probing. Off-the-shelf oscilloscopes can offer $256~GSa/s$ sampling rate and noise floor of hundreds of microvolts \cite{keysight}. Such equipment can be used to extract the power data.

We further analyzed the proposed attacking scheme after considering real-world artifacts, sampling rate and electrical noise. Algorithm \ref{alg:one} is based on extracting the total execution time of each tile. The execution can be easily distinguished from the idle state, and the convolution execution is much longer than the fully connected layer execution. Hence, we expect this timing analysis is robust even with artifacts. Algorithm \ref{alg:four} is an analysis of other extracted results without new physical signal measurements, so signal disturbance will not affect it. However, Algorithm \ref{alg:two} and Algorithm \ref{alg:three}, which are based on timing and power analysis, may be affected by the artifacts. Low sampling rate and high noise level may make it difficult to distinguish the ADC execution period from the analog computation period of the crossbar array. Timing analysis in Algorithm \ref{alg:two} will fail in this scenario. These conditions will also lead to an unreliable average power estimate, which may affect the power analysis in Algorithm \ref{alg:three}.

\begin{figure}[h]
  \centering
  \includegraphics[width=\linewidth]{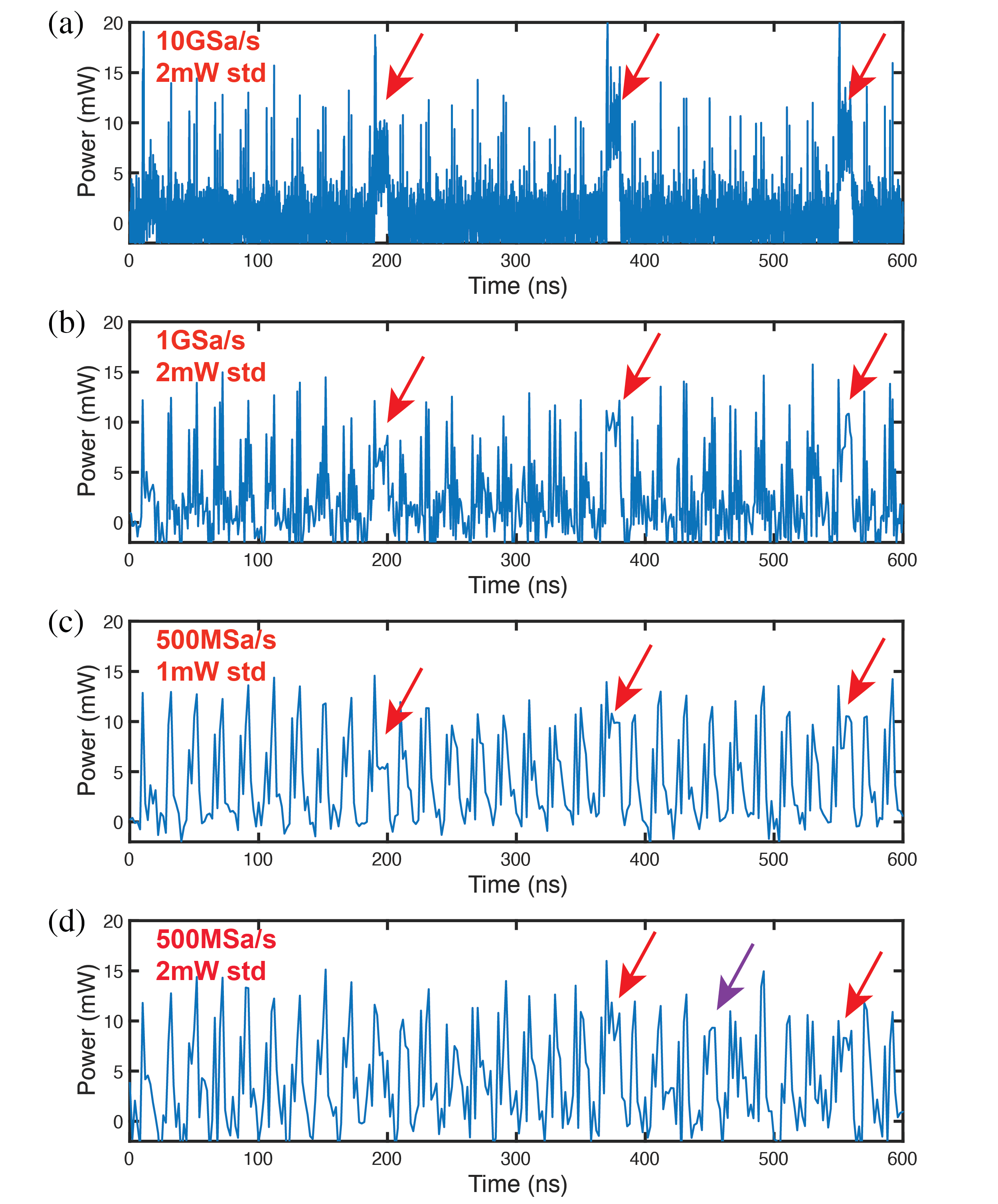}
  \caption{Power traces of the same tile for convolution execution with different sampling rates and noise levels. The sampling rates and noise standard deviation levels are labeled in (a) - (d).}
  \label{fig:12}
\end{figure}

To test the robustness of the proposed side-channel attack approach in real-word scenarios, we simulated power traces with different sampling rates and noise levels. In our initial simulation power traces without artifact injection are sampled at a rate of 10 GSa/s, and the analog operation and ADC execution periods exhibit significant differences, as shown in Figure \ref{fig:9}(b) and (c). In Figure \ref{fig:12}, the power traces of the same tile for convolution execution with different sampling rates and noise levels are displayed. Thanks to the time required to achieve a stable read output in the analog operation, the analog operation can still be identified even with these artifacts considered, as indicated by the red arrows. The ADC execution time can then be calculated from the intervals of these analog operations to execute Algorithm \ref{alg:two}. We noticed that at low sampling rate rates or high noise levels, the ADC and analog operation power signatures can become indistinguishable, as shown in Figure \ref{fig:12}(d). In Table \ref{tab:3}, we summarized the failure/success of Algorithm \ref{alg:two} at different artifact levels. We found the attack is more likely to fail at fully connected layers where VMMs are only executed once compared to repeated executions in convolution layers. Hence, there are fewer opportunities to identify the correct analog array execution period at fully connected layers.

\begin{table}[h]
  \centering
  \caption{{Algorithm 2 results with non-ideality.}}
  \label{tab:3}
    \begin{tabular}{|l|c|c|c|c|}\hline
        \diagbox[trim=l]
          {Sample\\Rate}{Noise \\Std} & 0 & 1~mW & 2~mW & 3~mW \\ \hline
        10 GSa/s & Success & Success & Success & Fail at FC \\    \hline
        1 GSa/s & Success & Success & Fail at FC & Fail at FC \\    \hline
        500 MSa/s & Success & Fail at FC & Fail & Fail \\    \hline
        200 MSa/s & Fail at FC & Fail & Fail & Fail \\    \hline
    \end{tabular}
\end{table}

\begin{figure}[h]
  \centering
  \includegraphics[width=\linewidth]{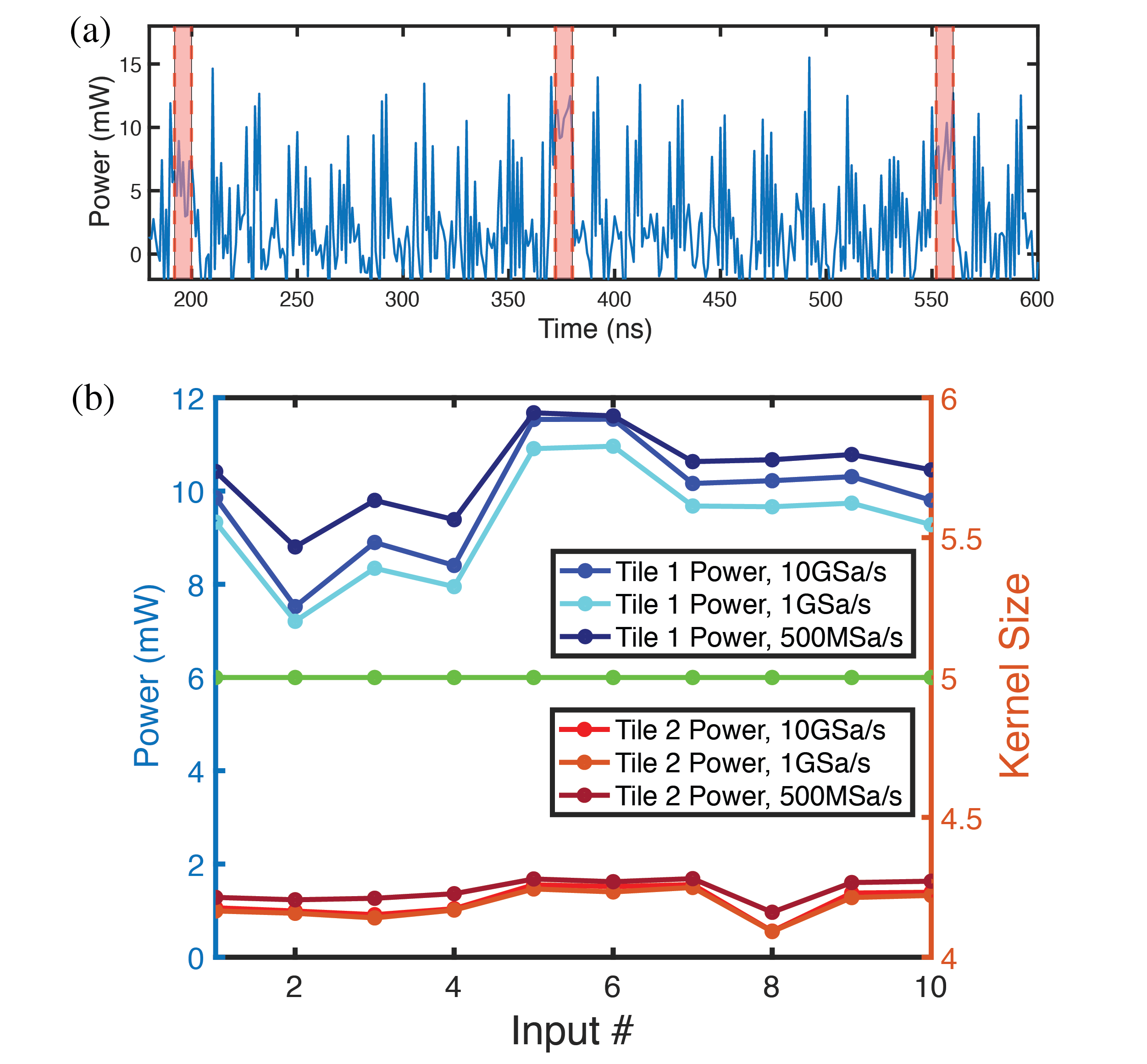}
  \caption{(a) Power traces for a convolution tile with 1GSa/s sampling rate and 2 mW noise standard deviation. The red region is the interested period for average power computation. (b) Average power of 2 convolution tiles and predicted kernel sizes with respect to different input images at 2 mW noise standard deviation and sampling rates from 500 MSa/s to 10 GSa/s.}
  \label{fig:13}
\end{figure}

The success of Algorithm \ref{alg:two} is the prerequisite for Algorithm \ref{alg:three}, since we need to identify the analog operation region before extracting its average power, as illustrated in the red regions of Figure \ref{fig:13}(a). We performed kernel size extraction tasks following Algorithm \ref{alg:three}, in the presence of artifacts injection. Even at noise levels of 2 mW standard deviation, the correct kernel size can still be extracted at sampling rates from 500 MSa/s to 10 GSa/s, as shown in Figure \ref{fig:13}(b). The failure/success of Algorithm \ref{alg:three} at different sampling rates and noise levels are summarized in Table \ref{tab:4}. An interesting result is when the sampling rate is reduced to 500 MSa/s, where it becomes difficult to identify the analog operation period, as shown in Figure \ref{fig:12}(d). However, if assuming the average power can still be calculated, the correct kernel size can still be obtained (Figure \ref{fig:13} (b)). These results prove the kernel size extraction method is robust against artifacts, likely due to the fact that effect of white noise can be effectively averaged out across multiple convolution executions.

\begin{table}[h]
  \centering
  \caption{{Algorithm 3 results with non-ideality.}}
  \label{tab:4}
    \begin{tabular}{|l|c|c|c|c|}\hline
        \diagbox[trim=l]
          {Sample\\Rate}{Noise\\Std} & 0 & 1~mW & 2~mW & 3~mW \\ \hline
        10 GSa/s & Success & Success & Success & Success \\    \hline
        1 GSa/s & Success & Success & Success & Success \\    \hline
        500 MSa/s & Success & Success & - & - \\    \hline
        200 MSa/s & Fail & - & - & - \\    \hline
    \end{tabular}
\end{table}

In summary, the timing and power side-channel attack methods are robust after considering real-world non-idealities, such as sampling rate and electrical noise. Table 3 and Table 4 summarize the required sampling rate for the proposed approaches to work well at different noise levels. The required specs can be offered by off-the-shelf measurement equipment \cite{keysight}. The required sampling rate can be further relaxed if the attacker can control the clock frequency of the system or effectively filter out noise from the measured signals.

\section{Countermeasures}

Techniques for preventing side-channel attack can be based on eliminating the correlation between the leaked information and the secret information \cite{kocher1999differential}, which is the DNN model in our case. For the proposed attacking algorithms, the countermeasures fall into three categories.

The first approach is to eliminate sensitive information leaked from timing analysis. For instance, scrambling the start time of the tiles or invoking dummy tiles with random delay times can obfuscate the layer sequence analysis. If the execution time of tiles are scrambled, we may find all tiles has the same start/end time and execution duration. This will make it impossible to differentiate the layer sequence and layer type. Inserting dummy input bits can increase the Conv layer execution latency. Dummy inputs can make it impractical to retrieve the output feature size from the execution time analysis, and make the iterative algorithm that searches the padding size ineffective. However, the above-mentioned techniques add penalties to both power consumption and latency of the system.

The second approach is to eliminate sensitive information leaked from ADC execution. The ADC execution time is directly related to the number of mapped columns in a tile and used to extract the layer size. Thus, one can add fake inputs to the ADCs in the partially mapped tile to match the ADC execution time as fully mapped tiles. Tiles will all have the same ADC execution time and the attack method based on ADC execution latency will fail. This approach will increase latency and the ADC power consumption.

The final approach is to mask the crossbar power consumption differences among tiles used in the same DNN layer, which will make the kernel size extraction algorithm ineffective. A potential technique is to add dummy conductance values to devices in unused columns. However, this method does not work if all tiles are column-wise fully mapped. Another approach is to devise other mapping schemes to balance and equalize the power consumption at analog computing. For example, thermal-aware weight mapping methods discussed in \cite{shin2020thermal} shows potential in balancing and reducing power consumption. Equalization techniques reduce the power leakage information by creating a consistent power side-channel profile. Equalization has been proved reliable to enhance the resistance of AES engines to power side-channel attacks \cite{wang2017power} \cite{cheng2022lightweight} \cite{tokunaga2009securing}. One can apply different weight mapping methods to different tiles to conceal the correlation between power consumption and mapped rows. However, this will introduce more workload in data pre-processing.

\section{Conclusion}
While mixed-signal IMC architectures may potentially assist with the obfuscation of sensitive data
by reducing the degree of data movement and limit memory access, we demonstrate how measurable electrical characteristics can still pose a security vulnerability. To perform reliable power trace analysis, we developed a dynamic power simulator based on a TSMC 28 nm process. We scrutinized the security vulnerability of analog IMC systems for DNN inference acceleration by proposing a series of side-channel attack analysis algorithms. Our analysis showed it is possible to uncover the complete model architecture information without any prior knowledge of the NN model. 

Using the proposed techniques, we were able to systematically uncover all layers in the example model and successfully reconstruct the full NN model. The proposed approach showed it is feasible to probe an IMC inference engine using algorithms that can work with other convolutional and dense DNN architectures. This study highlights the nature of security patches that may be required at the hardware abstraction, such as scrambling the timing information, adding dummy cycles to the ADC, and masking crossbar power using different weight mapping methods.

\ifCLASSOPTIONcompsoc
  \section*{Acknowledgments}
\else
  \section*{Acknowledgment}
\fi

This work was supported in part by SRC and DARPA through the Applications Driving Architectures (ADA) Research Center.

\ifCLASSOPTIONcaptionsoff
  \newpage
\fi

\bibliographystyle{IEEEtran}
\bibliography{ref}



\begin{IEEEbiography}[
{
\includegraphics[width=1in,height=1.25in,clip,keepaspectratio]{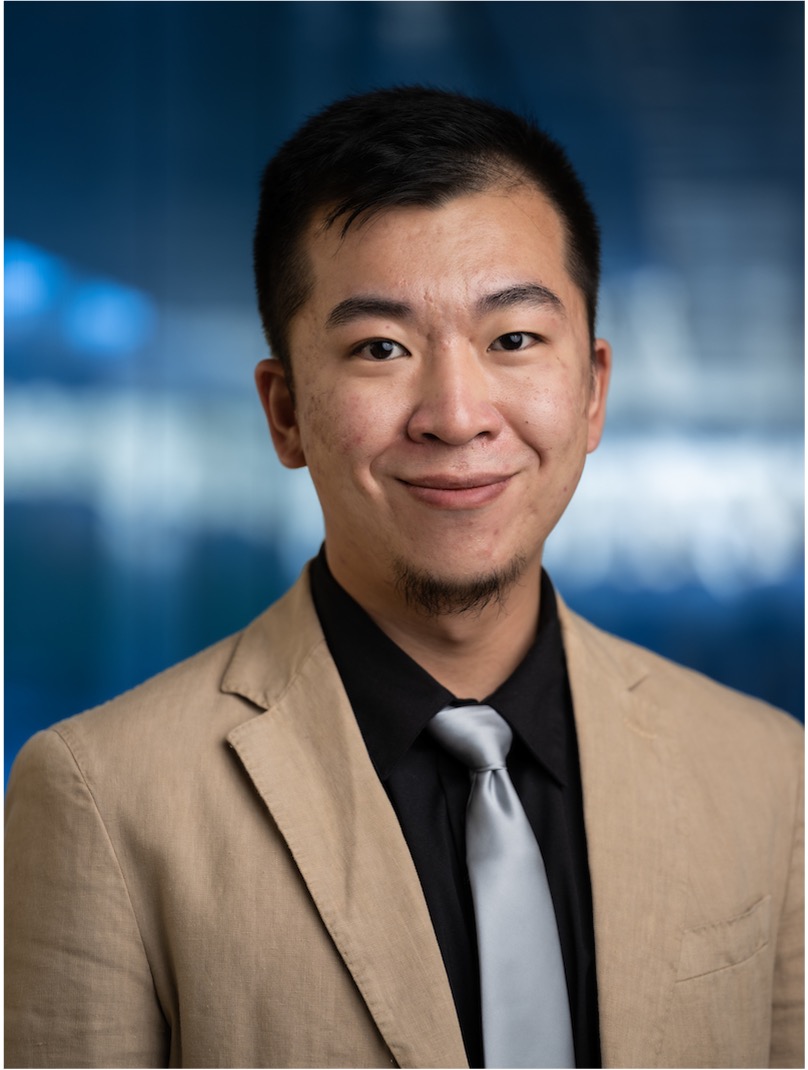}
}
]
{Ziyu Wang}
(Member, IEEE) received the B.E. degree from Tsinghua University, Beijing, China, in 2019, and the M.S. degree in electrical and computer engineering from the University of Michigan, Ann Arbor, MI, USA, in 2021. He is currently pursuing the Ph.D. degree in the Department of Electrical Engineering and Computer Science, the University of Michigan, Ann Arbor, MI, USA. His research interests focus on vulnerability analysis of emerging analog in-memory computing accelerator for deep neural network, as well as designing secure and reliable in-memory computing systems.
\end{IEEEbiography}

\begin{IEEEbiography}[
{
\includegraphics[width=1in,height=1.25in,clip,keepaspectratio]{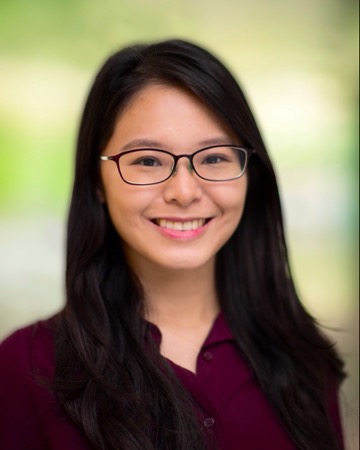}
}
]
{Fan-hsuan Meng}
received the B.S. degree in electrical engineering and the M.S. degree in electronics engineering from National Tsing Hua University, Hsinchu City, Republic of China, in 2014 and 2016, respectively. She worked as a process integration engineer for 7 nm FinFET devices in TSMC, Hsinchu City, Republic of China, from 2016 to 2017. She is currently pursuing the Ph.D. degree in the Department of Electrical Engineering and Computer Science, the University of Michigan, Ann Arbor, MI, USA. Her research interests include memristive devices, and its application for neuromorphic computing. She is currently working on system level optimization for in-memory computing based neural network accelerators.
\end{IEEEbiography}

\begin{IEEEbiography}[
{
\includegraphics[width=1in,height=1.25in,clip,keepaspectratio]{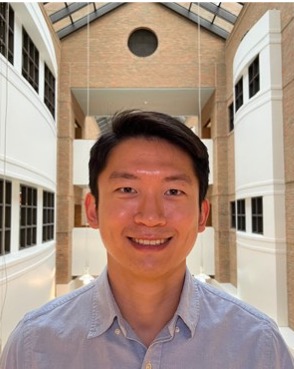}
}
]
{Yongmo Park}
(Graduate Student Member, IEEE) received the B.S. degree in electrical and electronic engineering from Yonsei University, Seoul, Korea, in 2019. He is currently pursuing the Ph.D degree with the Department of Electrical Engineering and Computer Science, the University of Michigan, Ann Arbor, MI, USA. His research interests include resistive-random access memory (RRAM) and analog compute-in-memory (CIM) systems for data intensive computations including deep neural networks. He was a recipient of the 2022 IBM PhD Fellowship Award Program. 
\end{IEEEbiography}

\begin{IEEEbiography}[
{
\includegraphics[width=1in,height=1.25in,clip,keepaspectratio]{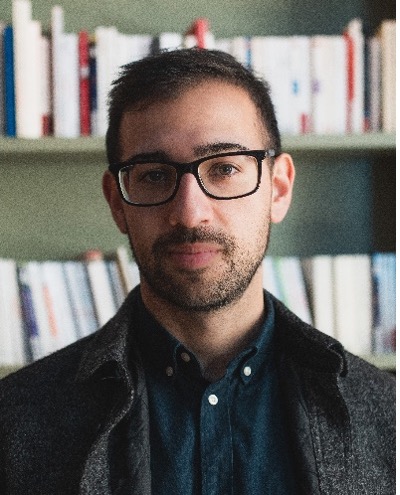}
}
]
{Jason K. Eshraghian}
(Member, IEEE)  received the Bachelor of Engineering (Electrical and Electronic)
and Bachelor of Laws degrees from The University of Western Australia, WA, Australia, in 2017, and
the Ph.D. degree from The University of Western Australia in 2019. From 2019 to 2022, he was a
Post-Doctoral Research Fellow at the University of Michigan, MI, USA. He is currently an Assistant
Professor with the Department of Electrical and Computer Engineering, University of California,
Santa Cruz. His current research interests include neuromorphic computing, resistive random access
memory (RRAM) circuits, and spiking neural networks. He was awarded the 2019 IEEE Very Large
Scale Integration Systems Best Paper Award, the Best Paper Award at the 2019 IEEE Artificial
Intelligence Circuits and Systems Conference, and the Best Live Demonstration Award at the 2020
IEEE International Conference on Electronics, Circuits and Systems.
\end{IEEEbiography}

\begin{IEEEbiography}[
{
\includegraphics[width=1in,height=1.25in,clip,keepaspectratio]{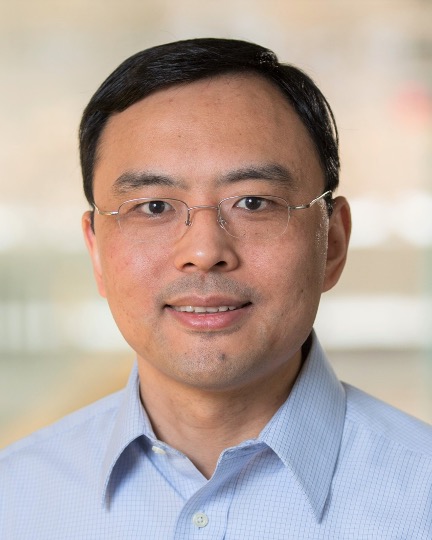}
}
]
{Wei D. Lu}
(Fellow, IEEE) received the B.S.
degree in physics from Tsinghua University, Beijing,
China, in 1996, and the Ph.D. degree in physics
from Rice University, Houston, TX, USA, in 2003.
From 2003 to 2005, he was a Post-Doctoral Research
Fellow at Harvard University, Cambridge, MA,
USA. He joined the Faculty of the University
of Michigan in 2005. He is currently a Professor with the Electrical Engineering and Computer
Science Department, University of Michigan. His
research interests include resistive-random access
memory (RRAM), memristor-based logic circuits, neuromorphic computing
systems, aggressively scaled transistor devices, and electrical transport in lowdimensional systems. He was a recipient of the NSF CAREER Award.
\end{IEEEbiography}
\end{document}